\documentclass[12pt,preprint]{aastex}

\def\Msun{$M_{\odot}$}

\def\Rsun{$R_{\odot}$}
\def\kms{kms$^{-1}$}
\def\vlsr{$V_{\rm LSR}$}

\def\h13cop{H$^{13}$CO$^+$}
\def\hc5n{HC$_{5}$N}

\def\t32{$J=3-2$}
\def\h13cn{H$^{13}$CN}
\def\cc34s{CC$^{34}$S}
\def\n2hp{N$_2$H$^+$}

\def\13co{$^{13}$CO}
\def\c18o{C$^{18}$O}
\def\ch3cn{CH$_{3}$CN}
\def\c34s{C$^{34}$S}
\def\3423{$3_4-2_3$}

\def\deg{\hbox{$^{\circ}$}}
\def\arcmin{\hbox{$^{\prime}$}}
\def\arcsec{\hbox{$^{\prime\prime}$}}

\def\cc{cm$^{-3}$}
\def\degree{$^{\circ}$}
\def\kms{km s$^{-1}$}

\def\vlsr{$V_{\rm LSR}$}

\def\h2{$H_{2}$}

\slugcomment{Accepted for publication in The Astrophysical Journal}

\shorttitle{IRC+10216's Innermost Envelope}
\shortauthors{Shinnaga et al.}

\begin{document}


\title{IRC+10216'S INNERMOST ENVELOPE -- THE ESMA'S VIEW}


\author{Hiroko Shinnaga\altaffilmark{1}, 
Ken H. Young\altaffilmark{2}, 
Remo P. J. Tilanus\altaffilmark{3,5},  
Richard Chamberlin\altaffilmark{1}, 
Mark A. Gurwell\altaffilmark{2}, 
David Wilner\altaffilmark{2}, 
A. Meredith Hughes\altaffilmark{2}, 
Hiroshige Yoshida\altaffilmark{1}, 
Ruisheng Peng\altaffilmark{1}, 
Brian Force\altaffilmark{1}, 
Per Friberg\altaffilmark{3},  
Sandrine Bottinelli \altaffilmark{4}, 
Ewine F. Van Dishoeck\altaffilmark{4}, and 
Thomas G. Phillips\altaffilmark{1}
} 
\email{shinnaga@submm.caltech.edu}

\altaffiltext{1}{California Institute of Technology Submillimeter Observatory (CSO)
   111 Nowelo St. Hilo HI 96720.}
\altaffiltext{2}{Harvard-Smithsonian Center for Astrophysics, 60 Garden Street, Cambridge, MA 02138.}
\altaffiltext{3}{Joint Astronomy Centre, 660 North A'ohoku Place, University Park, Hilo HI 96720}
\altaffiltext{4}{Leiden Observatory, Leiden University, P.O. Box 9513, NL-2300 RA Leiden, The Netherlands}
\altaffiltext{5}{Netherlands Organization for Scientific Research, P.O. Box 93138, NL-2509 AC The Hague, The Netherlands}


\begin{abstract}
We used the Extended Submillimeter Array (eSMA) in its most extended configuration 
to investigate the innermost (within a radius of $\sim$ 290 $R_{*}$ from the star) circumstellar envelope (CSE) 
of IRC+10216.  
We imaged the CSE using HCN and other molecular lines with a beam size of 0.\arcsec22 $\times$ 0.\arcsec46, 
deeply into the very inner edge ($\sim$ 15 $R_{*}$) of the envelope where the expansion velocity is only $\sim$ 3 \kms.  
The excitation mechanism of hot HCN and KCl maser lines is discussed.  
HCN maser components are spatially resolved for the first time on an astronomical object.  
We identified two discrete
regions in the envelope: a region with a radius of $\lesssim$ 15 $R_{*}$, where
molecular species have just formed and the gas has begun to be accelerated
(region I) and a shell region (region II) with a radius of ~23 $R_{*}$  and a
thickness of ~15 $R_{*}$, whose expansion velocity has reached up to 13 \kms,
nearly the terminal velocity of ~ 15 \kms. 
The Si$^{34}$S line detected in region I shows a large expansion velocity of 16 \kms~ due to strong wing components, 
indicating that the emission may arise from a shock region in the innermost envelope. 
In region II, the $P.A.$ of the most copious mass loss direction was found to be $\sim 120 \pm 10$\degree, which  
may correspond to the equatorial direction of the star.  Region II contains a torus-like feature.  
These two regions may have emerged due to significant differences in the size distributions of 
the dust particles 
in the two regions.  
\end{abstract}





\section{INTRODUCTION}\label{intro}
  IRC+10216 (Neugebauer \& Leighton 1969)  is the most studied carbon star.   
Because of proximity (distance of 135 pc (Le Bertre 1997)), 
it shows a very rich spectrum of 
molecular and atomic lines particularly at millimeter and submillimeter wavelengths (e.g., Groesbeck et al. 1994).  
As such, IRC+10216 is an important target for circumstellar and interstellar chemistry.  
The carbon star 
is a red giant that is evolving toward the post asymptotic giant branch (AGB) phase.  
It has a high mass-loss rate of up to 10$^{-5}$ \Msun yr$^{-1}$, and is therefore enshrouded by 
optically thick circumstellar shells.  

  Observing maser transitions of various molecules towards evolved stars is 
a powerful tool to study their innermost circumstellar envelopes, in particular the   
recent mass-loss activities and history of such stars.  
Silicon monoxide (SiO), water (H$_{2}$O) and hydroxyl (OH) 
maser transitions have been most frequently used to investigate the innermost 
mass-loss activities of 
oxygen rich evolved stars 
\citep{mor76,rei78,mor79,dia03,shi04}.  

  Hydrogen cyanide (HCN) is one of the most abundant molecular species in the 
circumstellar envelopes of carbon stars \citep{rid78} and 
is known to show maser action in various vibrational states 
(e.g., Ziurys \& Turner 1986 for $v \ge$ 1; Izumiura et al. 1995 for $v$=0).  
Since HCN is a linear triatomic molecule, it has three fundamental vibrational modes, ($v_{1}$, $v_{2}$, $v_{3}$); 
$v_{1}$ is the C--H stretching mode, $v_{2}$ is the doubly degenerate bending mode, 
and $v_{3}$ is the C--N stretching mode.  
Many observational studies have been done using HCN transitions in the circumstellar envelope of IRC+10216 and other carbon stars. 
Amazingly, some of the HCN maser lines are in very high energy levels (4000 K or above), 
far above than the temperature of the photosphere or the dust temperature of the region.  
The HCN component in the $v$=(0, 1$^{\rm 1f}$, 1) state, which is one of the high energy level transitions, 
is known to perturb with the $v$=(0, 4$^{0}$, 0) state  
through Coriolis interaction \citep{hoc67,cha71}, first empirically identified by Gebbie et al. (1964).  
The (0, 1$^{\rm 1f}$, 1) and (0, 4$^{0}$, 0) system have a high amplification in the 336.5 and 310.9 $\mu$m lines 
due to the fact that the vibrational transition moments are sharply peaked at $J=10$ \citep{lid67}.  
These rovibrational transitions  were detected towards IRC+10216 \citep{sch03}.  
This paper reports detections of two other lines at such high energy levels.   

\citet{luc92} used the Plateau de Bure interferometer to study the details of the inner circumstellar 
envelope of IRC+10216 with the $v$=(0, 2$^{0}$, 0) $J=2-1$ transition at 3 mm and they 
confirmed that the emission may be weak maser emission occurring in an acceleration region of the circumstellar envelope.
However, the components of the maser were not significantly resolved with their beam size of 2.4\arcsec $\times$ 1.1\arcsec.  

The Extended Submillimeter Array (eSMA) is a heterogeneous array that combines the signals 
detected with the eight 6-meter Submillimeter Array (SMA) antennas, 
the Caltech Submillimeter Observatory (CSO) 10.4 meter telescope, 
and the James Clerk Maxwell Telescope (JCMT) 15 meter telescope, operating at short millimeter and submillimeter wavelengths.  
Including the CSO in the interferometer, one can extend the longest baseline up to $\sim$782 meters, which 
improves the spatial resolution by $\sim$40 \%.    
The eSMA is one of the instruments that allows us to study the gas distribution and the kinematics of the innermost circumstellar envelope 
of this 
carbon star in detail where the acceleration of the gas is taking place.  
Here we report an observational study on the properties of the innermost circumstellar envelope 
of IRC+10216 using HCN and other molecular lines with the eSMA.  
We detected a few hot HCN lines including the one in the $v$=(0, 1$^{\rm 1f}$, 1) state 
mentioned in previous paragraph.  

\section{OBSERVATION AND DATA REDUCTION}\label{obs}
We observed IRC+10216 with HCN $J=3-2$ transitions in various vibrational states along with other molecular 
transitions 
with the eSMA \citep{bot08} on 2008 April 14 (UT).  
The total bandwidth of $\sim$ 4 GHz (2 GHz from each sideband, i.e., 264.357 $-$ 266.337 GHz and 274.357 $-$ 276.338 GHz) was covered 
with a frequency resolution of 812.5 kHz
\footnote{Note that JCMT's receiver 
had a more limited frequency coverage from 
264.770 to 266.338 GHz and from 274.357 to 275.927  GHz.  }. 
The data were taken under moderate weather conditions with  
zenith opacities of $\tau \sim$ 0.15 -- 0.18 measured at 225 GHz.  
Atmospheric opacities at the observing frequency were not measured 
during the observation, however, they are close to what we measured at 
225 GHz because these frequencies are close.  
Integration time for one data point was 30 seconds.  
The track lasted 5.5 hours, and about 4 hours of on-source data were acquired. 
The SMA was configured in its very extended array (maximum baselines to 508 m); the eSMA then covered baselines ranging from 25 to 782 m 
($\sim$ 20 to 692 k$\lambda$ or $\sim$ 0.\arcsec2 to 10\arcsec).  
The phase tracking center was set at 
RA = 9$^{\rm h}$ 47$^{\rm m}$ 57.$^{\rm s}$390, Decl. = 13\deg 16\arcmin 43.\arcsec898 (J2000). 
The synthesized beam size was 0.\arcsec 46 $\times$ 0.\arcsec 22 with a position angle of 39.9\degree.   

Data reduction and calibration were done using the MIR package\footnote{http://cfa-www.harvard.edu/$\sim$cqi/mircook.html} including passband and gain calibrations as well as continuum subtraction.  
IRC+10216 was observed together with quasars  3C273 and 1058+015.  
Passband calibration was done using the afore mentioned quasars.  

To make the eSMA observations, we used half wave plates (HWPs) at the JCMT and at the CSO 
to try to match the polarization angles of the receiver mixers with those at the SMA antennas.  
Based on an aperture efficiency of 75\% of the SMA antennas, the estimated aperture efficiencies 
of the JCMT and the CSO telescopes were found to be 40\% and 30\%, respectively.  
The CSO's aperture efficiency is typically $\ga$ 60\% when the telescope is used as a single-dish telescope.  
The degraded efficiency of the CSO telescope was due to a failure of the computer that controlled the HWP. 
The polarization angle of the CSO's 
mixer on the sky was not matched to the SMA's.  
The receiver temperatures of the SMA antennas and the CSO antenna ranged from $\sim$ 85 to 140 K, 
while the JCMT antenna's receiver temperature was $\sim$ 290 K.  
The double sideband system temperature of the SMA antennas and the CSO telescope were 
ranging between $\sim$ 200 -- 500 K, while the double sideband system temperature of the JCMT telescope was ranging 
between $\sim$ 300 -- 600 K.  

While the continuum emission of IRC+10216 is marginally resolved on the longest baselines (see Section \ref{continuum}), 
we have performed self-calibration of the visibility phase gains using the continuum, 
approximated by a point source, as our primary phase calibration step. 
Flux calibration was achieved by setting the continuum component of IRC+10216 to 1.18 Jy, consistent with well calibrated observations obtained in the SMA compact array in March 2008 whose beam size was about 3\arcsec. 
Based on the estimated flux of 1.18 Jy for IRC+10216, 
the calculated flux of the quasar 1058+015 was 2.53 Jy, which was 9 \% higher than the flux in the calibrator catalog\footnote{http://sma1.sma.hawaii.edu/callist/callist.html?data=1058\%2B015} measured on the same date (2008 April 17).  
Since the quasar is linearly polarized by $\sim$ 25 -- 30 \% at this frequency, 
the 9\% difference may be due to polarization.    
Note that, by applying self-calibration using the continuum component of IRC+10216,  
the data taken on the quasar 1058+015 showed  
that atmospheric decorrelation was eliminated significantly even on long baselines.  


Image processing including clean was done using Astronomical Image Processing System (AIPS\footnote{http://www.aips.nrao.edu/}) 
with natural weighting of the visibilities. 
Natural weighting was chosen to achieve a high signal-to-noise ratio for detecting and imaging weak components along 
with strong ones.  
%
Comparing the data taken with the SMA with the data taken with the eSMA, 
it was found that the inclusion of the CSO and JCMT antennas, i.e. the eSMA, improved 
the signal-to-noise ratio on the longest baselines by ~40\% compared to that of the SMA alone.

In order to estimate missing flux of the two HCN transitions in the $v$=0 and  (0, 1$^{\rm 1e}$, 0) states, 
we obtained a single-dish spectrum of the two HCN transitions 
with the JCMT on 2008 May 11th (UT).  
The spectrum was acquired at an elevation near 30\degree.  
The main beam efficiency was measured to be 65\% with a beam size of $\sim$ 19\arcsec~ 
at 266 GHz, by using the continuum emission of Saturn.  
We used 106.5 K for the main beam temperature of the planet. 
Note that the deep PH$_{3}$ absorption feature \citep{wei94} of the planet was taken into account 
when estimating the main beam temperature of the planet at 266 GHz.  
Conversion from the $T_{\rm A}^{*}$ scale to flux density scale was achieved with a conversion factor of 
32 Jy K$^{-1}$ for the JCMT data.  





\section{RESULTS AND DISCUSSION}\label{results}
\subsection{Continuum Component from Circumstellar Dust and the Stellar Photosphere}\label{continuum}
\begin{figure}
\epsscale{.60}
\plotone{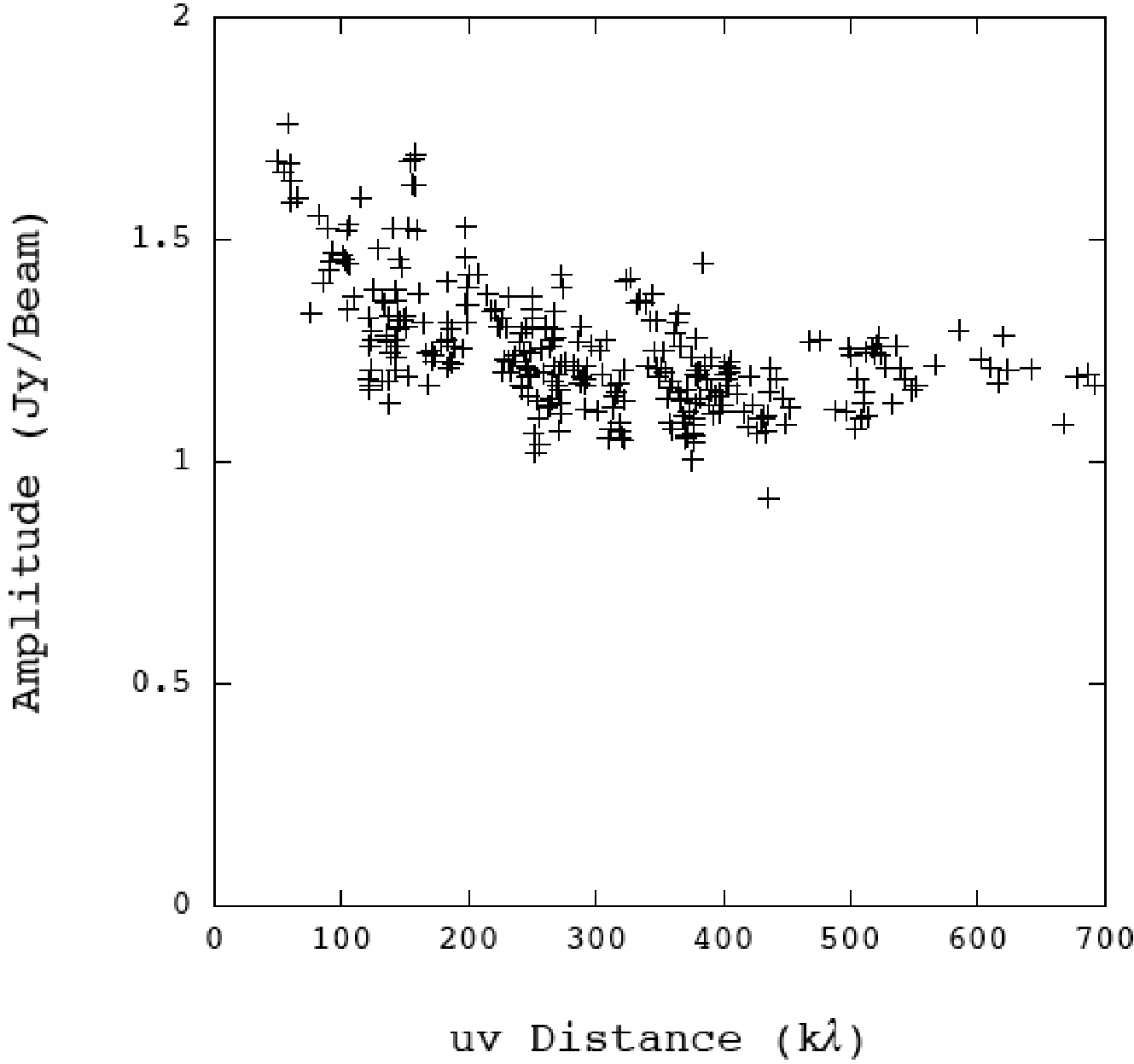}
\caption{Visibility amplitude versus  uv distance of the continuum emission of the circumstellar dust and photosphere at 1.1 mm.  
\label{f1}}
\end{figure}

Continuum data were created from the portions of the spectra which appear to be free from line emission.  
Figure \ref{f1} shows visibility amplitude versus  uv distance of the continuum component originating from 
the circumstellar dust, and photosphere of the central star, at $\sim$ 1.1 mm.  
The data points were averaged over every $\sim$ 5 k$\lambda$.  
The falloff of visibility amplitude with  uv distance demonstrates that the continuum component is 
spatially extended and partially resolved.   
In Figure 1, one can find that the amplitude is flattened at 400 k$\lambda$.  This means that 
there are components contributing to the continuum emission which 
are smaller than 27 $R_{*}$ or 62 AU in size.  
On the baselines shorter than 400 k$\lambda$, the amplitude of the continuum emission 
rises  as going towards shorter baseline length, which means that there 
are continuum components which we resolved whose radii $\ga$ 27 $R_{*}$.  
$R_{*}$ is estimated to be $\sim$ 500 \Rsun = 3.5 $\times$ 10$^{13}$ cm \citep{men01}.  

It is known that the continuum emission of IRC+10216 varies with a period of 635 days at submillimeter  as well 
as optical wavelengths \citep{jen02}.  
At 850$\mu$m, the average flux is 8.8 Jy and it changes by $\pm$ 0.95 Jy due to the pulsation \citep{jen02}.  
Taking the sinusoidal curve of Figure 7 in \citet{jen02}, at the time when 
the object was observed in 2008 March and April, the object is expected to be in the maximum phase (phase of 0.015 
by counting from the maximum phase). 
\citet{gro97} measured that the star's continuum emission extends up to 50\arcsec~ at the 3$\sigma$ level of their measurements at 1.3 mm.  
The continuum emission measured from our JCMT observation was 3.7 Jy at 1.1 mm, which is consistent with measurements done by other authors (e.g., Dehaes et al. 2007).  
\citet{sah89} estimated the continuum emission from the stellar photosphere 
would be 0.55 Jy at 1 mm.  It suggests that 
about half of the total continuum emission imaged with the eSMA may be due to the emission of the stellar photosphere, 
considering the measured flux of the continuum component with the SMA was 1.18 Jy (see Section 2).  
The rest of the flux measured with the eSMA may come from circumstellar dust components and 
low-level line emission which were not individually detected.  

\subsection{HCN and Other Molecular Species Detected with the eSMA}\label{lines}
\begin{figure}
\includegraphics[angle=0,scale=.70]{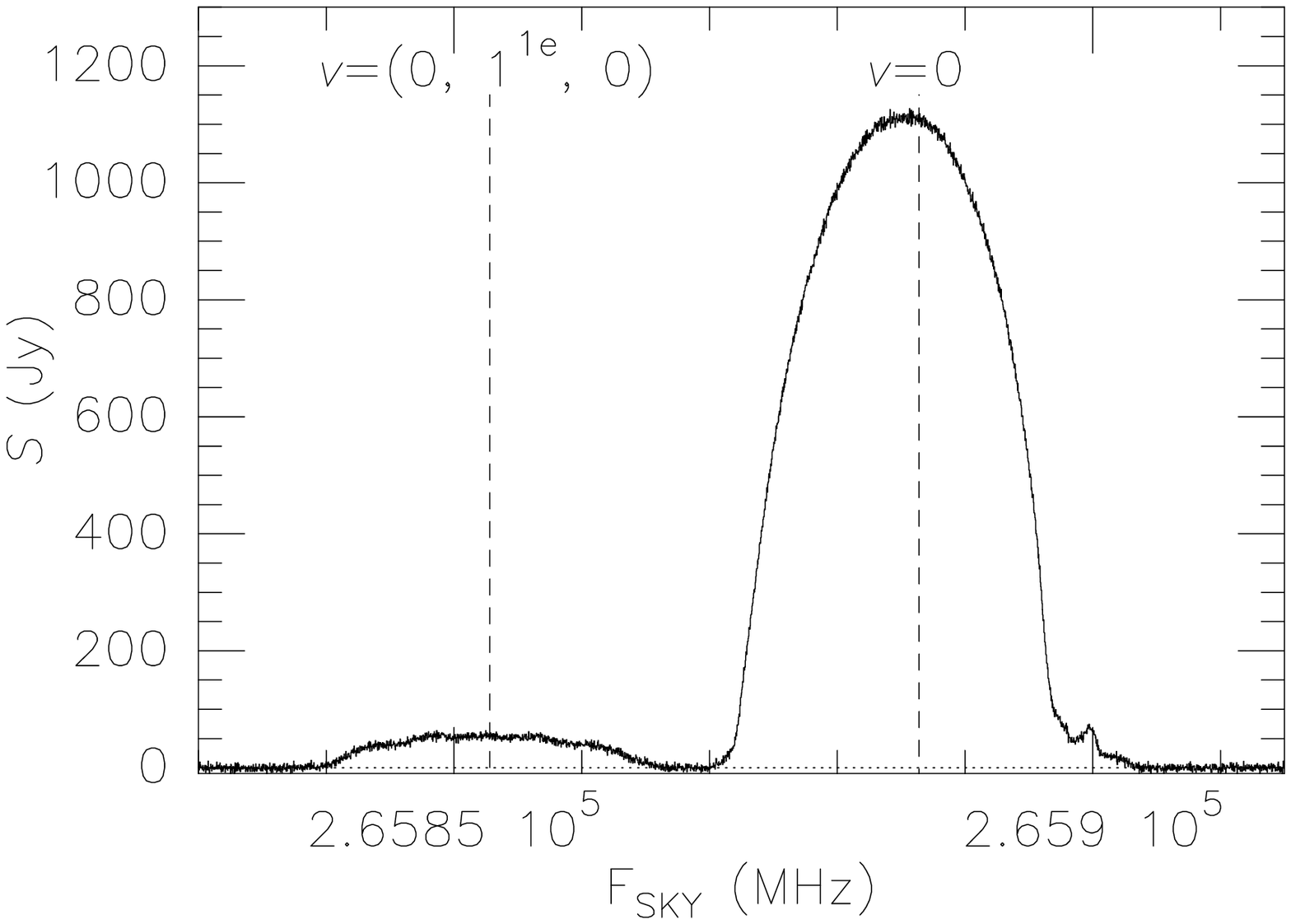}\\
\includegraphics[angle=0,scale=.70]{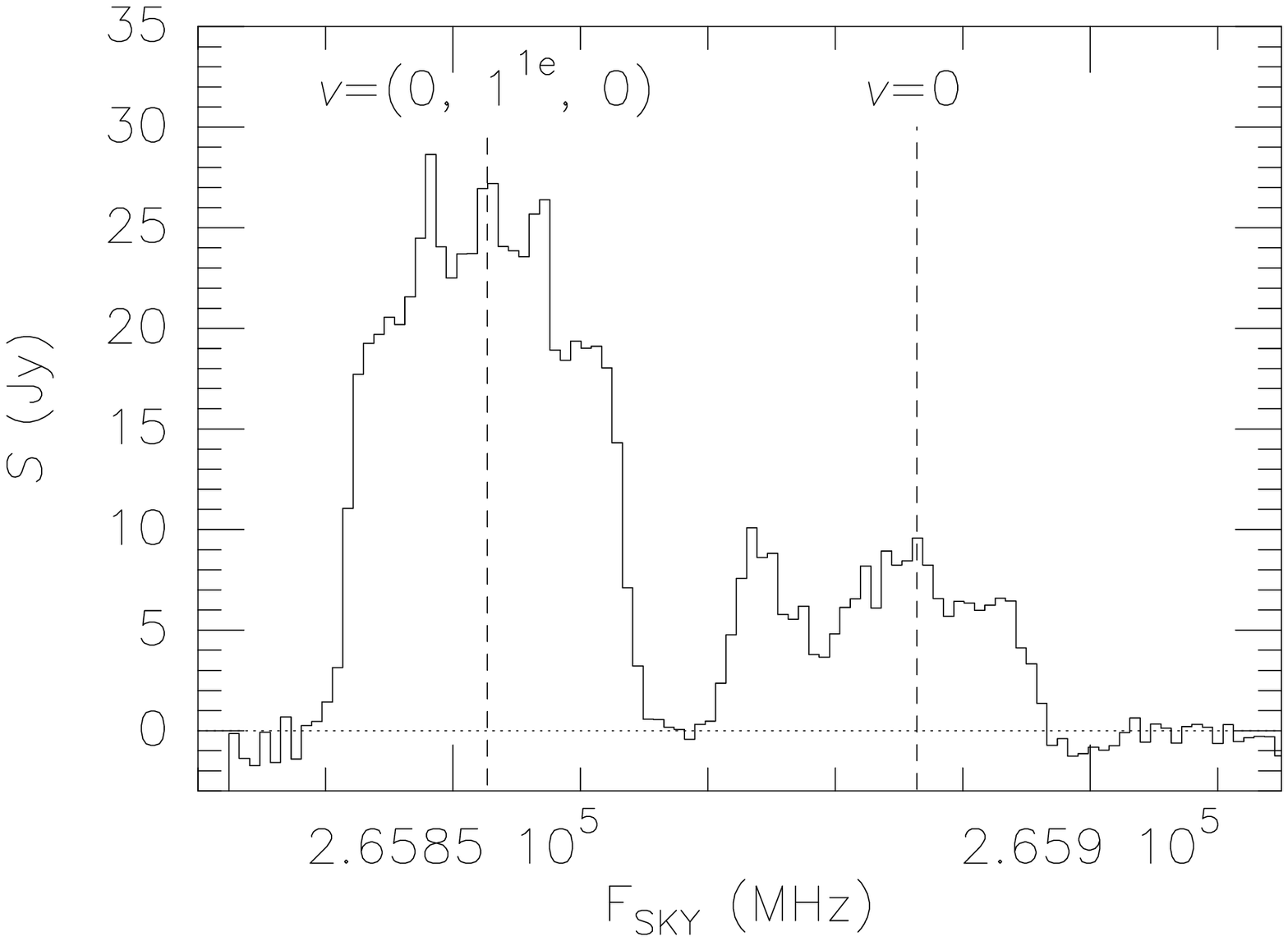}
\caption{HCN $J=3-2$ spectrum in the ground vibrational state and 
$v$=(0, 1$^{\rm 1e}$, 0) state taken at the JCMT (top) and with the eSMA (bottom).  
\label{f2}}
\end{figure}

\begin{figure}
\includegraphics[angle=0,scale=.70]{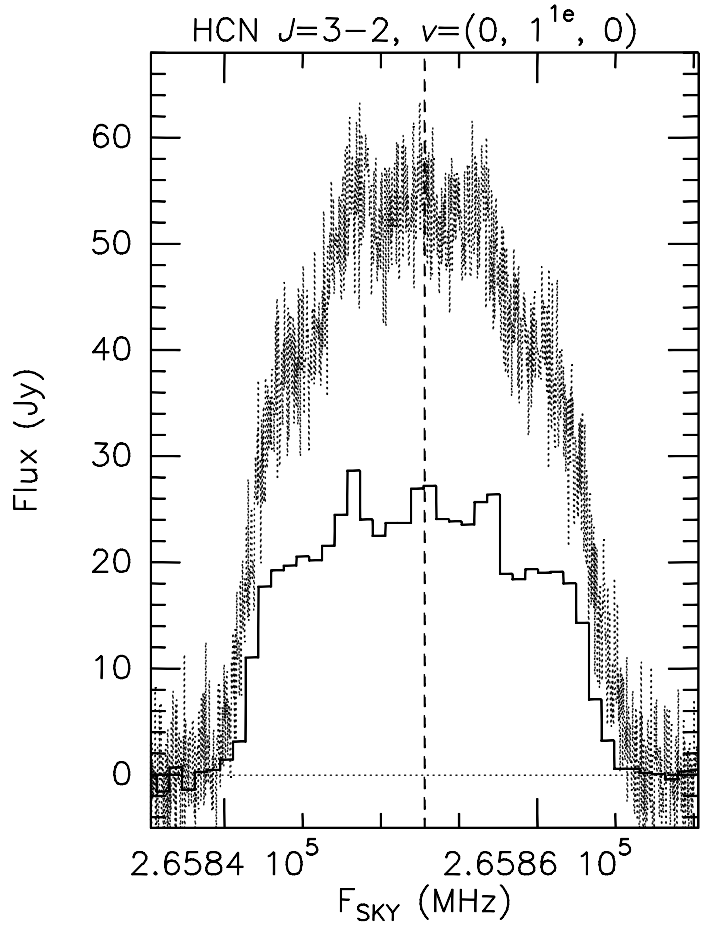}
\caption{Comparison of HCN $J=3-2$ $v=$(0, 1$^{\rm 1e}$, 0) spectra taken with the JCMT (dashed line) and with the eSMA (solid line).  
\label{f3}}
\end{figure}

Figure \ref{f2} 
shows the spectra of HCN $J=3-2$ transitions in the 
$v$=0 and (0, 1$^{\rm 1e}$, 0) states over sky frequency ($F_{\rm SKY}$) taken at the JCMT and at the eSMA, respectively.  
Figure \ref{f3} shows 
the comparison of the $J=3-2$, $v=$(0, 1$^{\rm 1e}$, 0) lines observed with the JCMT and the eSMA more closely.  
The eSMA spectra shown in Figure \ref{f2} and \ref{f3} were produced from a spectral cube generated by the image processing.  
To create the spectrum, a region was specified which includes all emission of the transitions detected above 3$\sigma$ level 
in the total integrated intensity map of the transitions.  

\citet{ziu86} observed HCN $J=3-2$, $v$=(0, 1$^{\rm 1e}$, 0) transition in 1986 March at the Kitt Peak 12m telescope.  
The stellar phase in 1986 March was near minimum.  
The peak flux observed in 2008 April was $\sim$ 25 \% higher than that observed in 1986.  
The increased flux might be due to variability of the stellar luminosity.  

While both the HCN $J=3-2$, $v$=0 and (0, 1$^{\rm 1e}$, 0) single-dish spectra 
show a typical parabolic-
shape profile that traces the expanding envelope, the line profiles of the transitions 
taken with the interferometer have strange irregular shapes with a few peaky components.  
The fractions of the flux in the $v$=0 and (0, 1$^{\rm 1e}$, 0) states detected by the eSMA are  
only 1 and 45 \% of the emission detected by the JCMT.  
Please note that the size of the JCMT's beam corresponds to 1.1  $\times$ 10$^{3}$  $R_{*}$ 
(= 2.6 $\times$ 10$^{3}$ AU).  
The rest, i.e., the  missing components, probably originating in the expanding envelope, 
are resolved out with this array configuration.  
The size of the missing components is unknown 
because mapping observation using a single-dish telescope has not done.  

For the line data sets, we discuss only the detected compact components.  
Physical properties estimated from the velocity widths follow in Sections \ref{distribution} and \ref{linewidth}.

In addition to the two HCN transitions above, seven other molecular transitions are clearly detected (Figure 4).  
One of the detected HCN transitions is in the vibrational state of (0, 1$^{\rm 1f}$, 1), mentioned in Section 1.  
Molecular lines were identified mainly by comparing the observed frequencies with 
the line catalogs provided by NIST (SLAIM), JPL and Universit$\ddot{\rm a}$t zu K$\ddot{\rm o}$ln (CDMS).  
When there were multiple candidates for a spectral line feature, 
we checked the expected line strengths of the different transitions 
over the observed frequency ranges and identified most likely one.  
For the lines that don't have other transitions in the band, we selected most likely one 
considering expected abundance in the circumstellar region.  

Table \ref{tbl-1} summarizes features of nine lines 
detected with the eSMA.  
It includes an isotopologue of silicon monosulfide, Si$^{34}$S ($v$=0), potassium chloride, KCl ($v=2$), 
and two other HCN transitions in 
different vibrational states, $v$=(1, 1$^{\rm 1f}$, 0) and (0, 1$^{\rm 1f}$, 1). 
The excitation temperatures of these transitions are about 5800 K and 4000 K, respectively.  
The HCN $v=0$ line has a relatively large shift to the red 
($\sim$ 2.7 MHz or 3 \kms).  
This may be due to absorption from the hyperfine transitions \citep{mor85} 
because of the nitrogen's quadrapole coupling.  
The rest of the lines have smaller differences between the catalog and observed frequencies ($\lesssim$ 700 kHz). 
The central velocities of molecular transitions may be slightly shifted because 
(1) some of the circumstellar shells are not perfect shells, 
(2) evolution of the molecular species, (3) optical depth of 
the molecular lines, and so on.  
The excitation of the HCN transitions will be discussed in Section \ref{excitation}. 
The distributions of these hot HCN transitions will be described in Section \ref{distribution}.  

We detected the $J=35-34$ transition of KCl in the $v=2$ state.  
KCl was first detected at 3 mm by \citet{cer87} along with other metals in the circumstellar envelope of IRC+10216.  
The transitions they detected were all in the $v=0$ state.  
This is the first detection of a KCl transition in a vibrationally excited state. 
The Si$^{34}$S $J=15-14$ transition has a broader line width (full width at half maximum of $\sim$ 9.2 \kms) with two peaks.  
It seems to have a wing component on the red-shifted side.   
The broad line feature of  Si$^{34}$S might be due to blending from other molecular line(s).
However, the lines near the frequency were not found in the available catalogs.  
The line shapes of these transitions will be discussed in Section \ref{linewidth}.  
The maps of these molecular transitions 
are shown in Section \ref{distribution}.  

\begin{figure}
\epsscale{.80}
\includegraphics[angle=0,scale=.60]{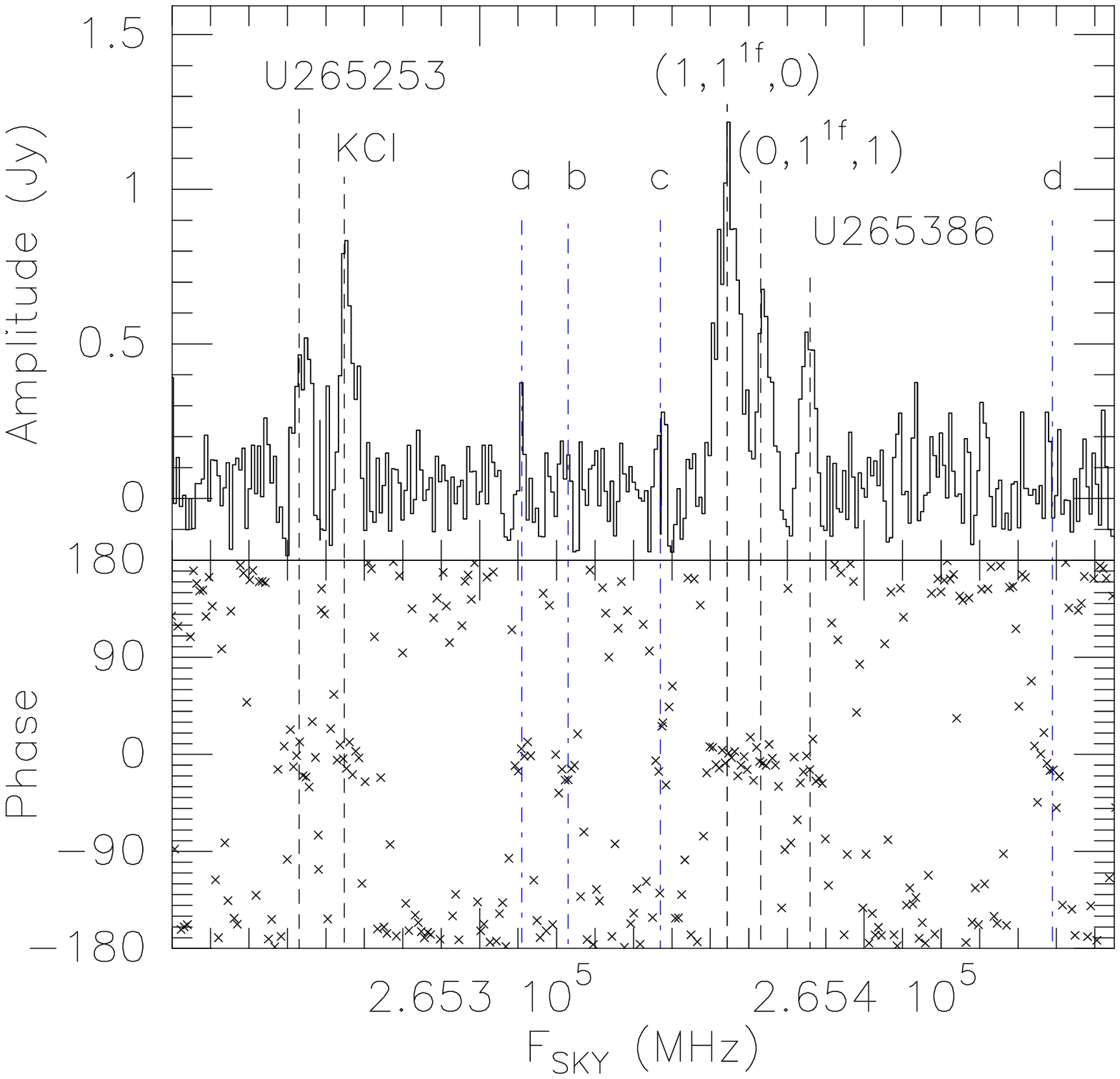}\\
\includegraphics[angle=0,scale=.60]{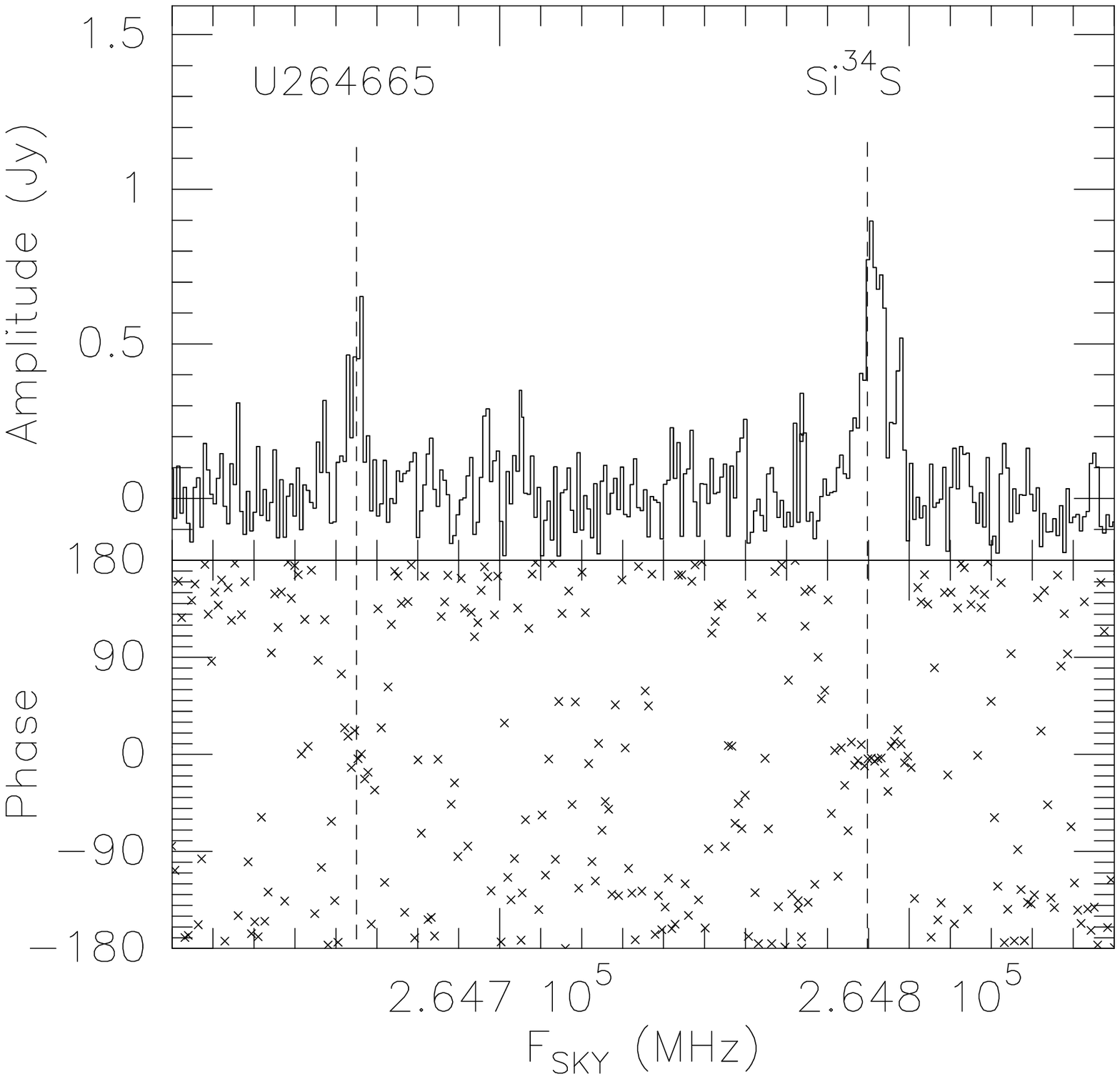}
\caption{Averaged spectra over all baselines.  
\label{f4}}
\end{figure}

There were four other spectral features whose phases are tightly clustered within $\sim$ 50\deg~ from zero 
over $\ga$ 6 -- 10 \kms, signifying higher likelihood of detection, albeit at much lower amplitudes compared 
to the nine molecular lines described above.  
Their frequencies of 265.311, 265.323, 265.347, and 265.449 GHz, 
are marked as a, b, c and d in Figure \ref{f4}.  
These four features are likely detections.  

\subsection{Excitation of the Hot HCN Lines and the KCl Lines}\label{excitation}
Excitation of the HCN lines in three different vibrational states and 
the KCl line are of interest because 
they exist in the innermost circumstellar envelope where dust and molecular species are forming and the molecular gas is being accelerated (see Section \ref{linewidth}).   

The excitation temperatures of the lower energy levels of these transitions, $\sim$ 5800 K for the $v$=(1, 1$^{\rm 1f}$, 0) state  
and $\sim$ 4000 K for the $v$=(0, 1$^{\rm 1f}$, 1) state, 
are much higher than the temperature at the inner edge of the envelope 
(600 -- 1700 K; e.g., Men'shchikov et al. 2001 and references therein).  
It would be difficult to excite the transitions collisionally unless very high density gas ($\ga$ 10$^{10}$ cm$^{-3}$ or so) is present in a large region ($\ga$ 5\arcsec $\sim$ 700 AU or $\sim$ 300 $R_{*}$).  
It's very unlikely that the condition is achieved for this case.  
The upper level of the molecules may be populated directly through the strong radiation of the stellar photosphere.  

\begin{figure}
\epsscale{.80}
\includegraphics[angle=0,scale=0.6]{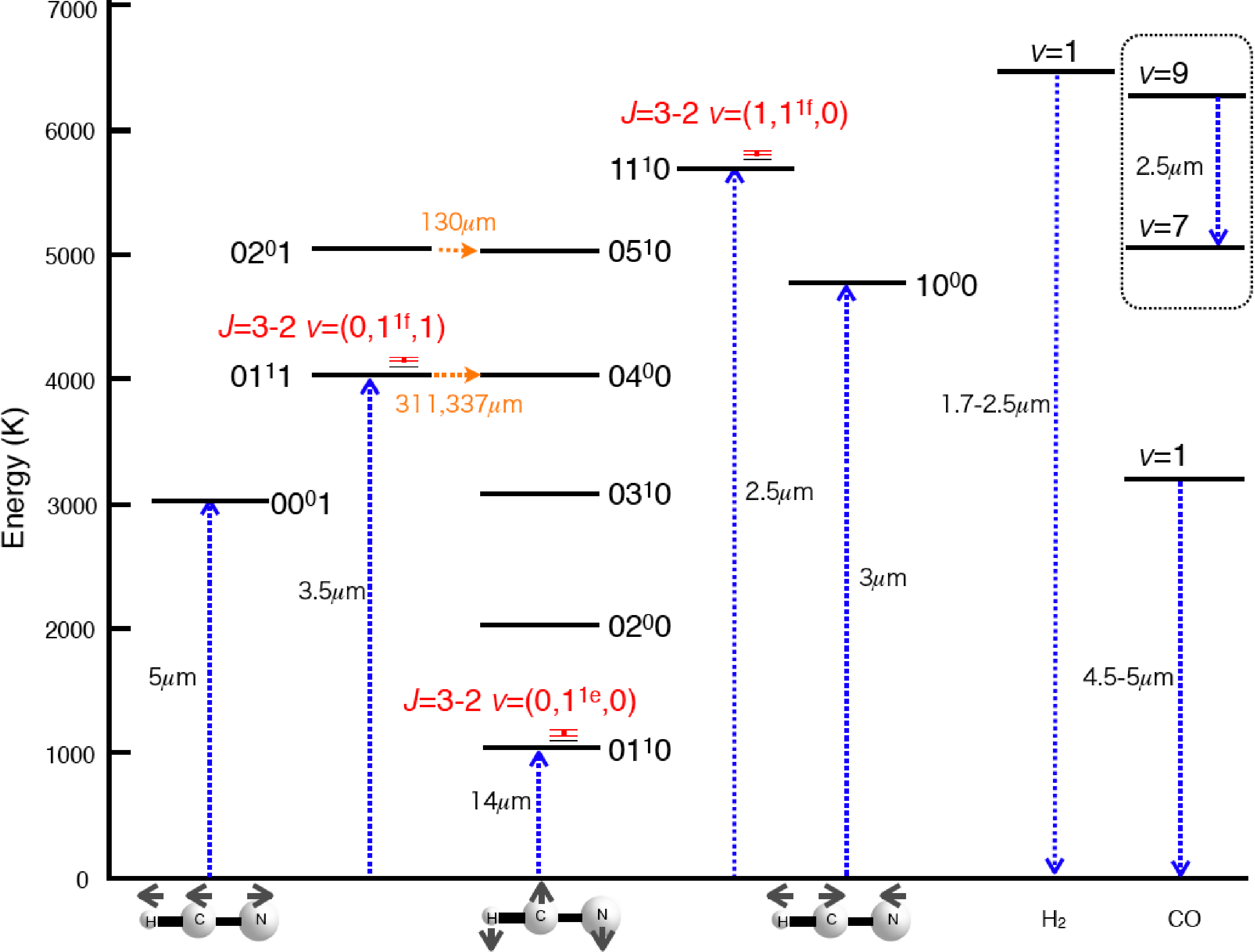}
\caption{Energy diagram of selected transitions and vibrational states of HCN, H$_{2}$, and CO.  
The red dots plotted between rotational states marked with red color indicate 
the observed HCN transitions with the eSMA.  
Under three vibrational modes (see Section \ref{intro}), three cartoons are drawn for 
each vibrational mode \citep{pic74}.  
The blue dotted thick arrows pointing upwards and downwards represent excitation and de-excitation of the transitions.  
The orange dotted arrows represent rovibrational maser transitions.  
The $v$=9-7 transition of CO encased with dotted line at top right is not drawn in the scale of the diagram.  
\label{f5}}
\end{figure}

Figure \ref{f5} illustrates energy diagrams of selected transitions of HCN, H$_{2}$, and CO.  
The observed rotational lines in three different vibrational states are drawn by thin horizontal lines and 
are magnified in the diagram.  
We propose that some of the HCN lines may be pumped through interaction with photons of some transitions of molecular hydrogen, those of CO lines, and the stellar radiation.  
The stellar luminosity of IRC+10216 is $\sim$ 2 $\times$ 10$^{4}$ L$_{\odot}$ and   
the peak of the spectral energy distribution (SED) of the object is near 10 $\mu$m (e.g. Cernicharo et al. 1999).  
In the near infrared band which is near the peak of the SED, 
such as at 2.5, 3.5, and 14 $\mu$m, 
the molecules may be pumped to the vibrational states of HCN in the $v$=(1, 1$^{1f}$, 0), (0, 1$^{1f}$.1), (0, 1$^{1e}$, 0), (1, 0$^{0}$, 0), and (0, 0$^{0}$, 1) states 
from the ground level to the exited states via the stellar radiation.  
Figure \ref{f5} also shows 
the excitation temperatures of 
H$_{2}~ v=$ 1 state are close to that of the HCN (1, 1$^{1}$, 0) state and 
some of the 
transitions have energy levels very similar to that of HCN (1, 1$^{1}$, 0) state, 
indicating that a direct and rapid photon exchange can occur among the transitions. 
Furthermore, since CO $v= 9-7$ transition is at $\sim$ 2.5 $\mu$m, such photon exchange can occur with the HCN $v=(1, 1^{1}, 0)$ state as well.  


The same excitation method, namely  
direct pumping from the strong infrared radiation at $\sim$14 $\mu$m of the stellar photosphere, 
can explain the excitation of the KCl transition in the $v=2$ state.   


These hot HCN lines and the KCl line may be maser lines as the stellar radiation must excite the transitions.  
If so, the maser lines of the transitions may not be fully saturated because their flux is not high.  

\subsection{Distribution of HCN, SiS, and KCl --- the Properties of the Innermost Circumstellar Envelope}\label{distribution}
\begin{figure}
\epsscale{.80}
\includegraphics[angle=0,scale=0.6]{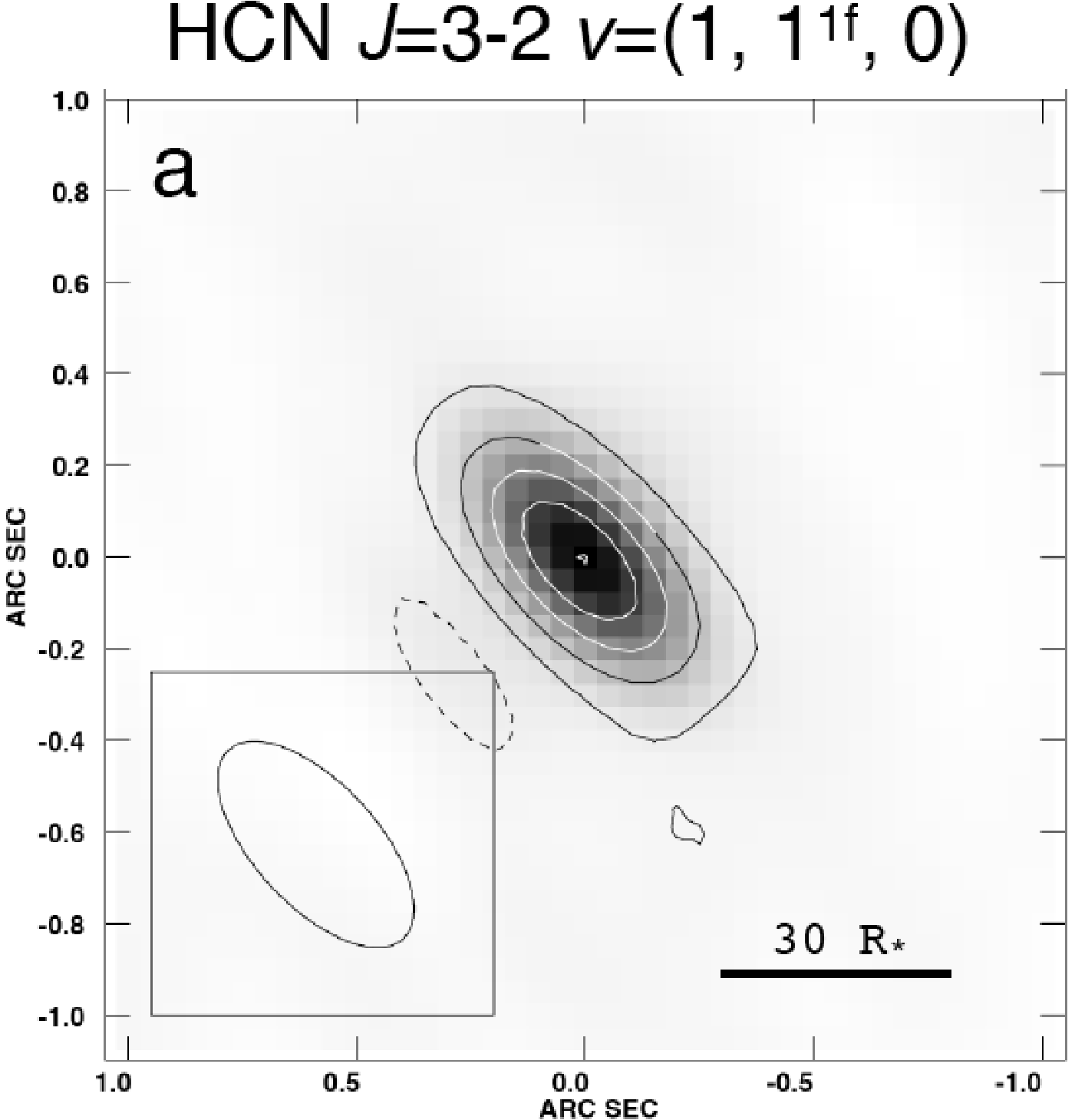}
\includegraphics[angle=0,scale=0.6]{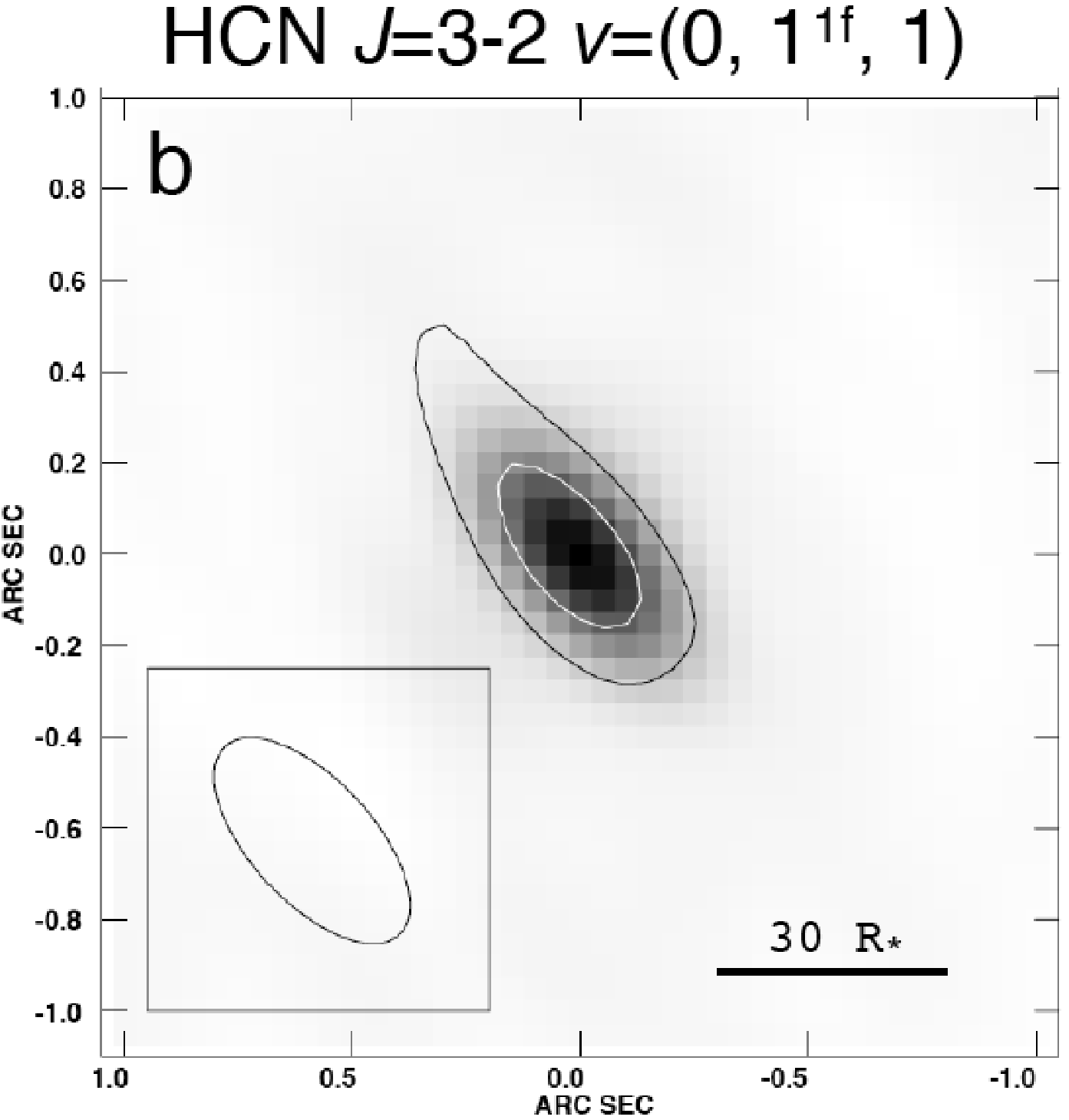}
\includegraphics[angle=0,scale=0.6]{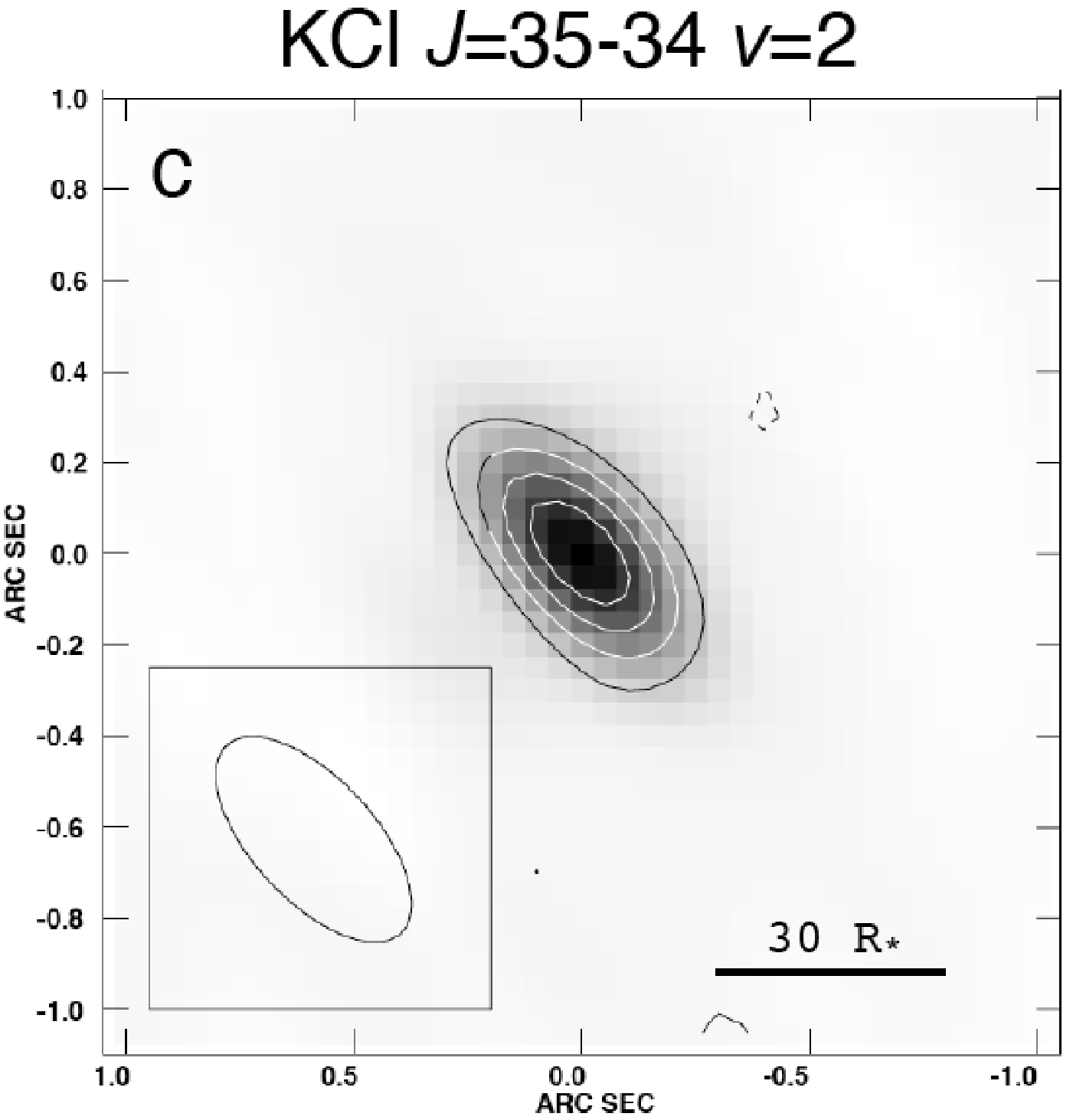}
\includegraphics[angle=0,scale=0.6]{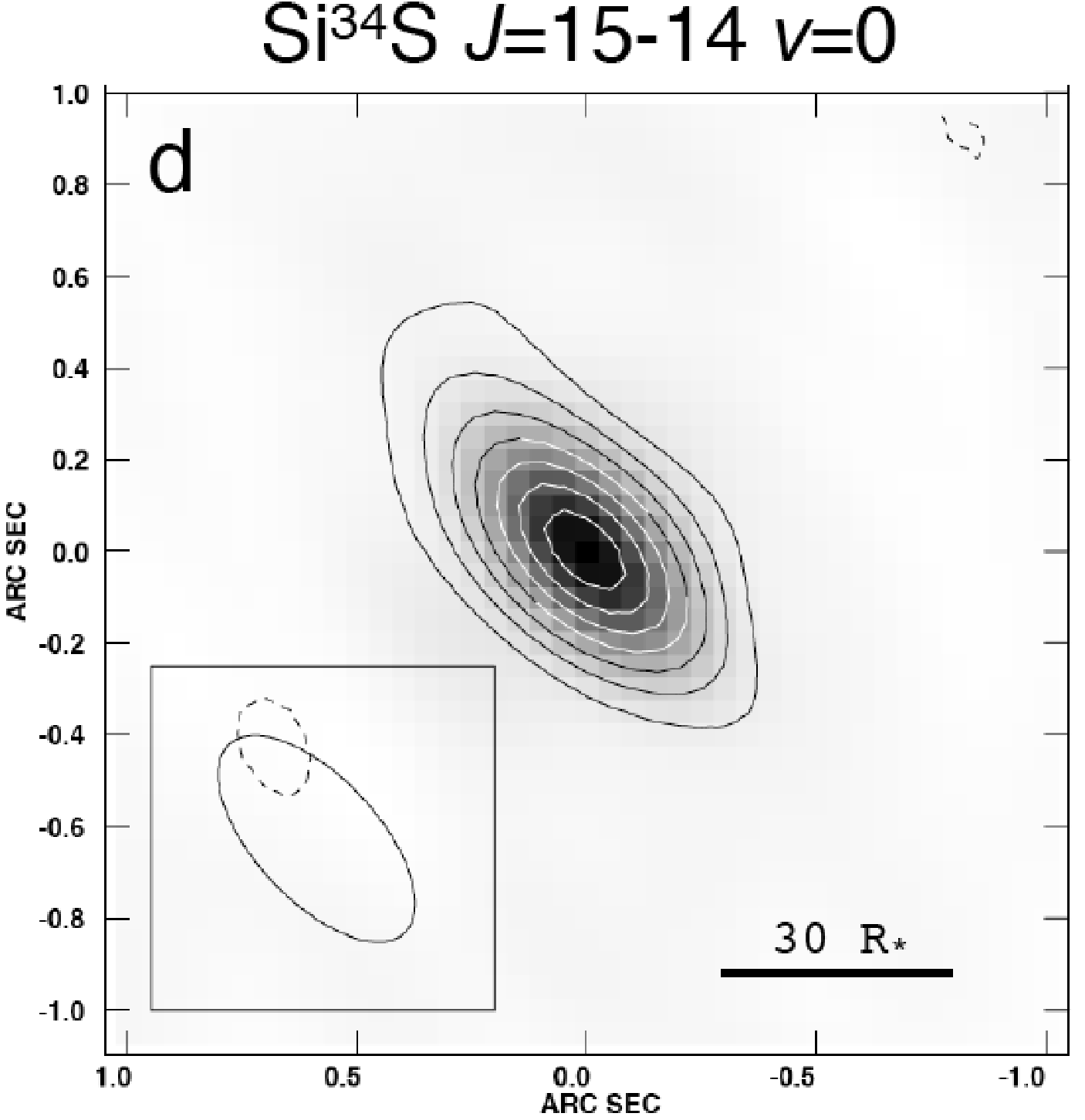}
\caption{Integrated intensity maps of (a) HCN $J=3-2$ transitions in the $v=(1, 1^{\rm 1f}, 0)$ state, 
(b) in the $v=(0, 1^{\rm 1f}, 1)$ state, 
(c) KCl $J=35-34$ in the $v=2$ state, (d) Si$^{34}$S $J=15-14$ in the $v=0$ state
overlaid on the continuum emission of the circumstellar dust and photosphere.  
Contours are set at -3, 3, 9, and 15 $\sigma$ (1 $\sigma$ $\sim$ 23 mJy).   
\label{f6}}
\end{figure}

Integrated intensity maps of hot HCN $J=3-2$ emission in the $v$=(1, 1$^{\rm 1f}$, 0) and $v$=(0, 1$^{\rm 1f}$, 1) states are 
shown in Figure \ref{f6} (a) and (b) respectively.  The distribution of the emission is not spatially resolved.  
The peak of the emission coincides with the stellar position.  
The radius of the distribution is less than 15 $R_{*}$.  
The compact distribution supports the scenario that the pumping arises 
close to the stellar photosphere or at inner edge of the circumstellar envelope. 

Figure \ref{f6} (c) and (d) show the integrated intensity maps of 
the transitions of KCl and Si$^{34}$S.  
Both maps show that the emission 
originates from a region very close to the central star, very similar to the HCN maps of the $v$=(1, 1$^{\rm 1f}$, 0) and $v$=(0, 1$^{\rm 1f}$, 1) states.  
Both of the KCl component and the Si$^{34}$S component are not significantly resolved.  
The Si$^{34}$S emission is partially resolved.  However, most of the emission originates  
from a compact region near the central star.  
The compact distribution of the Si$^{34}$S $J=15-14$ $v=0$ transition and the peaky line features 
(Figure \ref{f4}) may indicate that part of the emission arises from maser action.  
The distribution of KCl $v=2$ emission is very compact and 
the emission is not spatially resolved, most likely tracing inner edge of the envelope. 

\begin{figure}
\epsscale{.80}
\includegraphics[angle=0,scale=0.36]{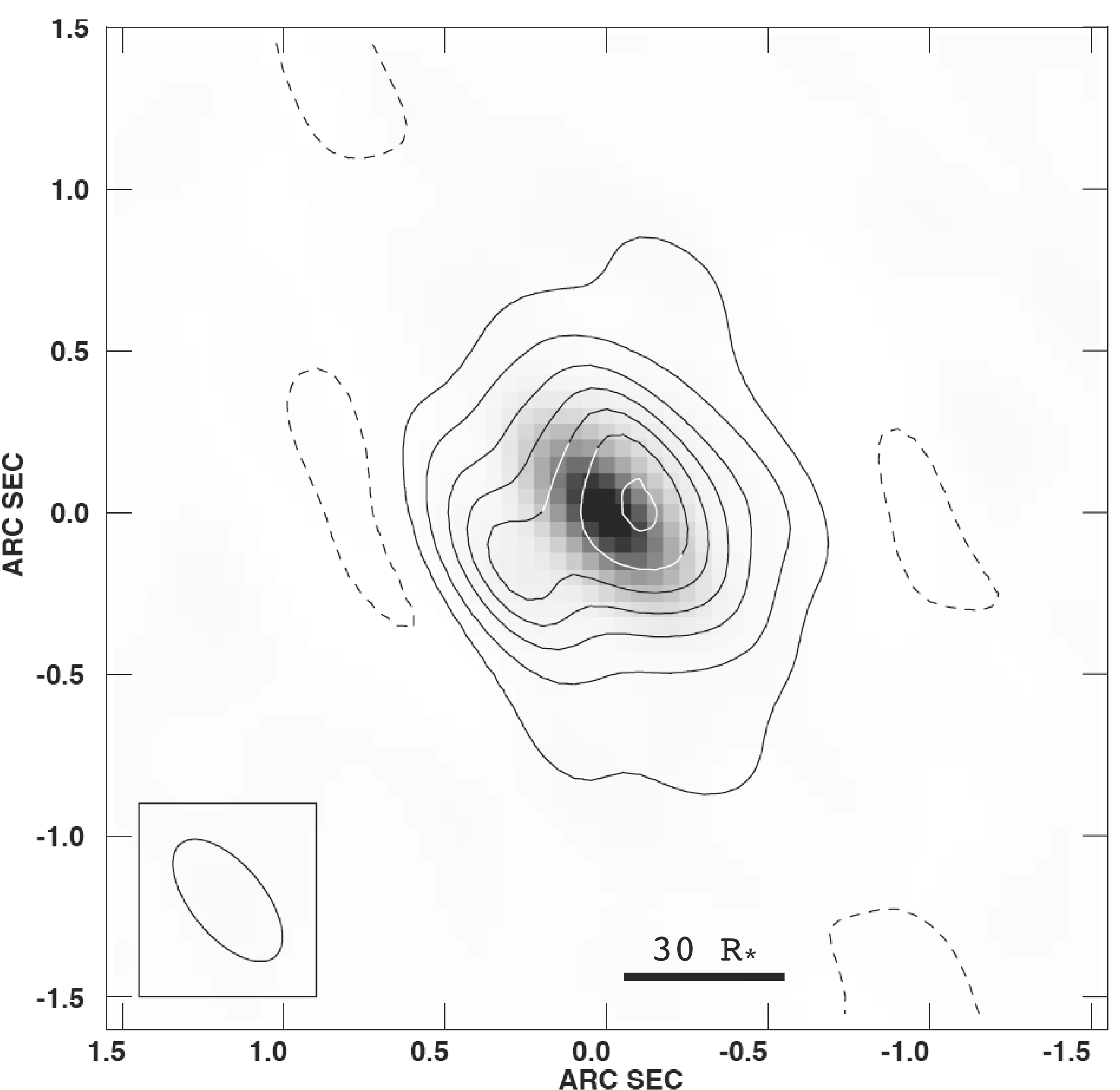}
\includegraphics[angle=0,scale=0.36]{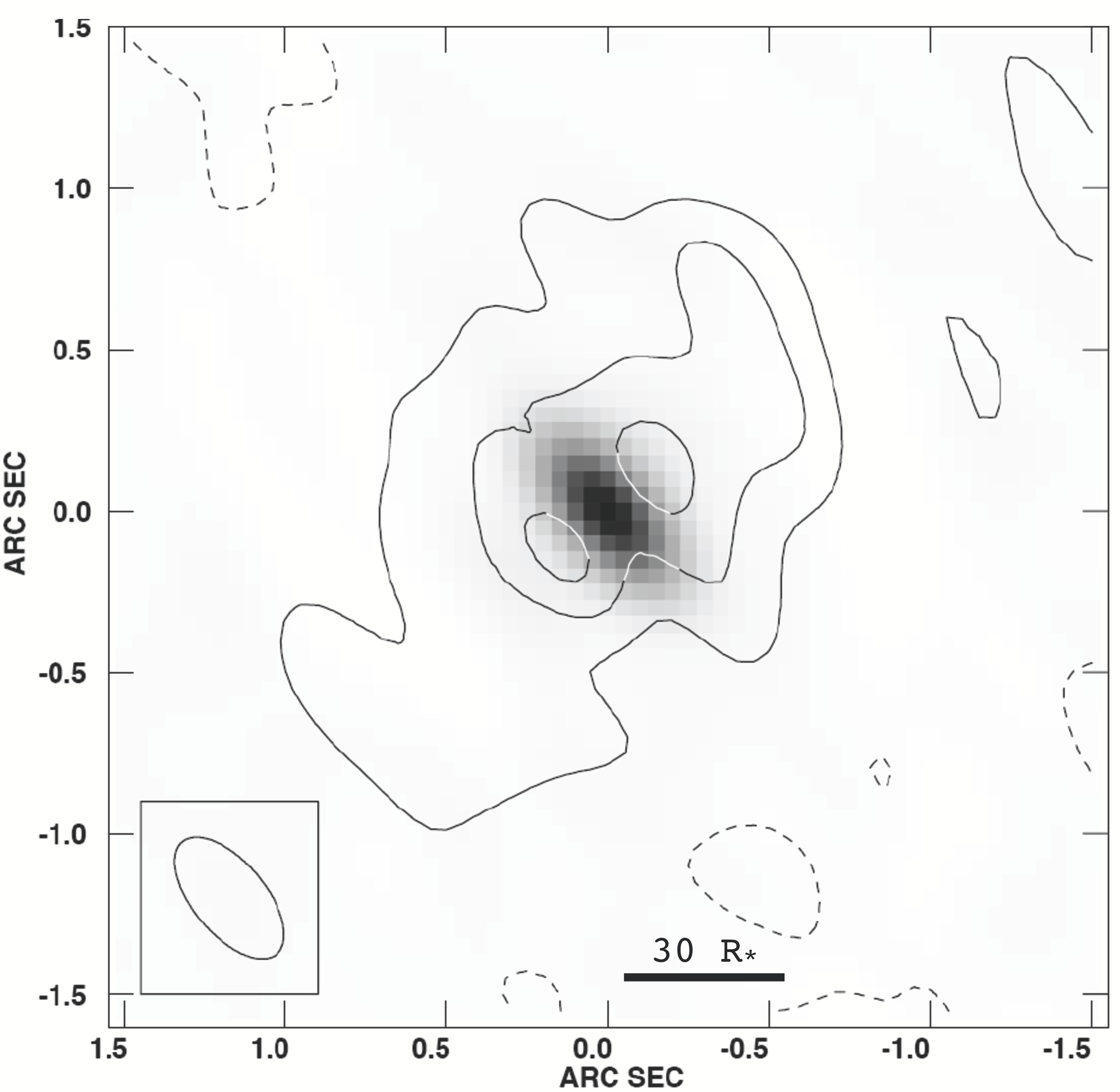}
\caption{Total integrated intensity maps (contours) of HCN $J=3-2$ $v=(0, 1^{\rm 1e}, 0)$ (left) and 
$v=0$ (right) states, overlaid on the continuum emission of the circumstellar dust and photosphere.  
For the $v=(0, 1^{\rm 1e}, 0)$ state map, the contours are drawn at 
-3, 3, 10, 20, 30, 40, 50, and  60 $\sigma$ (1 $\sigma$ = 57 mJy), respectively.   
For the ground vibrational state map, the contours are drawn at -3, 3, 6, and 9 $\sigma$ (1 $\sigma$ = 39 mJy).  
\label{f7}}
\end{figure}

Total integrated intensity maps of the HCN $v$=(0, 1$^{\rm 1e}$, 0) and $v=0$ transitions are shown in Figure \ref{f7}.  
In contrast to the maps shown in Figure \ref{f6} that are not spatially resolved, 
the emission is much more extended.  
Most of the emission is found within a radius of 1\arcsec, which corresponds to $\sim$ 60 $R_{*}$.  
One can find two peaks along the northwest - southeast direction on both maps.  
The two peaks of both maps are displaced from 
the stellar position. 
The position angles ($P. A.$) of the line drawing between two peak components in the $v$=(0, 1$^{\rm 1e}$, 0) and in the $v=0$ states are  
115\deg~ and 130\deg, respectively.  
The distances of the two condensations of the $v$=(0, 1$^{\rm 1e}$, 0) map are 90 mas and 300 mas from the central star.  
The two condensations of $v=0$ map are distributed more or less symmetrically at a distance of 200 mas from the central star.  
In the region at a distance of 200 mas from the star, 
the gas temperature is estimated to be a few 100 K (e.g., Keady et al. 1988).  

\citet{mur05} reported that there are features with a low degree of linear polarization identified at 1.5 $\mu$m in continuum emission.  
The authors concluded that low degree of linear polarization found inside the two features 
indicates that the density in the two regions is  higher than that of the surrounding region.  
The two features called NW ellipse and SE fan by \citet{mur05} may be the same entities as   
the two HCN peaks traced with the HCN $J=3-2$ transition in the $v$=(0, 1$^{\rm 1e}$, 0) state, 
because of their similarity in  shapes, 
P.A., and the fact that both of them trace high density regions.  
However, comparing the locations of the polarization minima reported by \citet{mur05} and the HCN $v$=(0, 1$^{\rm 1e}$, 0) peaks  in Figure \ref{f7} (left) relative to the position of the central star, 
the HCN peaks are located at a distance from the central star only 60\% of the distance between the polarization minima from the star.  
Please note that \citet{mur05} estimated the location of the central star based on their polarization map.  
It is not clear why the HCN peaks and the near infrared polarization minima lie on different locations.  
If it were due to time evolution, the two HCN peaks and the polarization minima might be tracing different entities 
because the HCN peaks are located closer compared to the polarization minima.  
It could be that the HCN peaks detected this time in 2008 might not have existed clearly when 
Murakawa et al made the observation in 2003 January.  
However, in that case, one can say that the direction of the most copious mass loss hasn't changed over about 5 years.  

From the two HCN peaks and those polarization features,  
we propose that this direction along the two HCN peaks of the HCN transitions may correspond to the direction of most 
conspicuous mass loss, and that this direction 
is the equatorial direction of the star.  
This proposition is consistent with the analysis of \citet{mur05} who discussed that the two linear polarization features may indicate the formation of an edge-on dust torus surrounding the carbon star. 
Discussion on the kinematics of the two HCN peaks and other surrounding components will follow in next section. 

Note that there are a few other high angular resolution observations at near infrared 
(e.g., Weigelt et al. 2002; Menut et al. 2007).  
However, the location of the central star in their maps is uncertain so that it is difficult to compare their observed features with ours.   
In particular, we can't find any of likely components that may correspond to the two HCN peaks/ the two polarization minima.  

At near infrared (2.2$\mu$m), \citet{kas94} and \citet{ski98} identified a 10\arcsec-size bipolar reflection nebula 
that is produced by scattering of radiation from the star by dust particles in the region. 
The axis of the bipolar reflection nebula is at $P. A.$ of $\sim$ 30\deg, which is  
almost perpendicular to that of the most conspicuous mass-loss direction described above.   
The axis of the bipolar reflection nebula supports the scenario that the mass loss is most copious in the direction of the two HCN peaks 
shown in the maps in Figure \ref{f7} and  that infrared radiation can pass 
preferentially 
in the direction perpendicular to the most conspicuous mass-loss direction.  

The significant difference in distribution between HCN emission in $v$= 0/(0, 1$^{\rm 1e}$, 0) states and 
the HCN emission in the $v$=(1, 1$^{\rm 1f}$, 0)/$v$=(0, 1$^{\rm 1f}$, 1) states or KCl and Si$^{34}$S emission 
leads us to conclude that 
%
there are two discrete regions in the innermost circumstellar envelope.  
One is a region of radius less than 15 $R_{*}$ where various molecular species have just formed and 
the gas is not accelerated by stellar radiation yet.   We name this region region I.  
The other is a region outside of the region I, region II, where spectral lines indicate higher expansion velocity.  
These two discrete regions may have been created because 
the size distributions of the condensed dust particles are significantly different in these regions (see also Section \ref{linewidth}).  
Dust nucleation may have further proceeded because temperature decreases rapidly as going outwards from the central star.  
We will discuss the properties of the region II in next section. 

\subsection{The Kinematics and The Distributions of the HCN $v$=(0, 1$^{\rm 1e}$, 0) Maser Clumps}\label{kinematics}
\begin{figure}
\epsscale{.80}
\includegraphics[angle=0,scale=0.5]{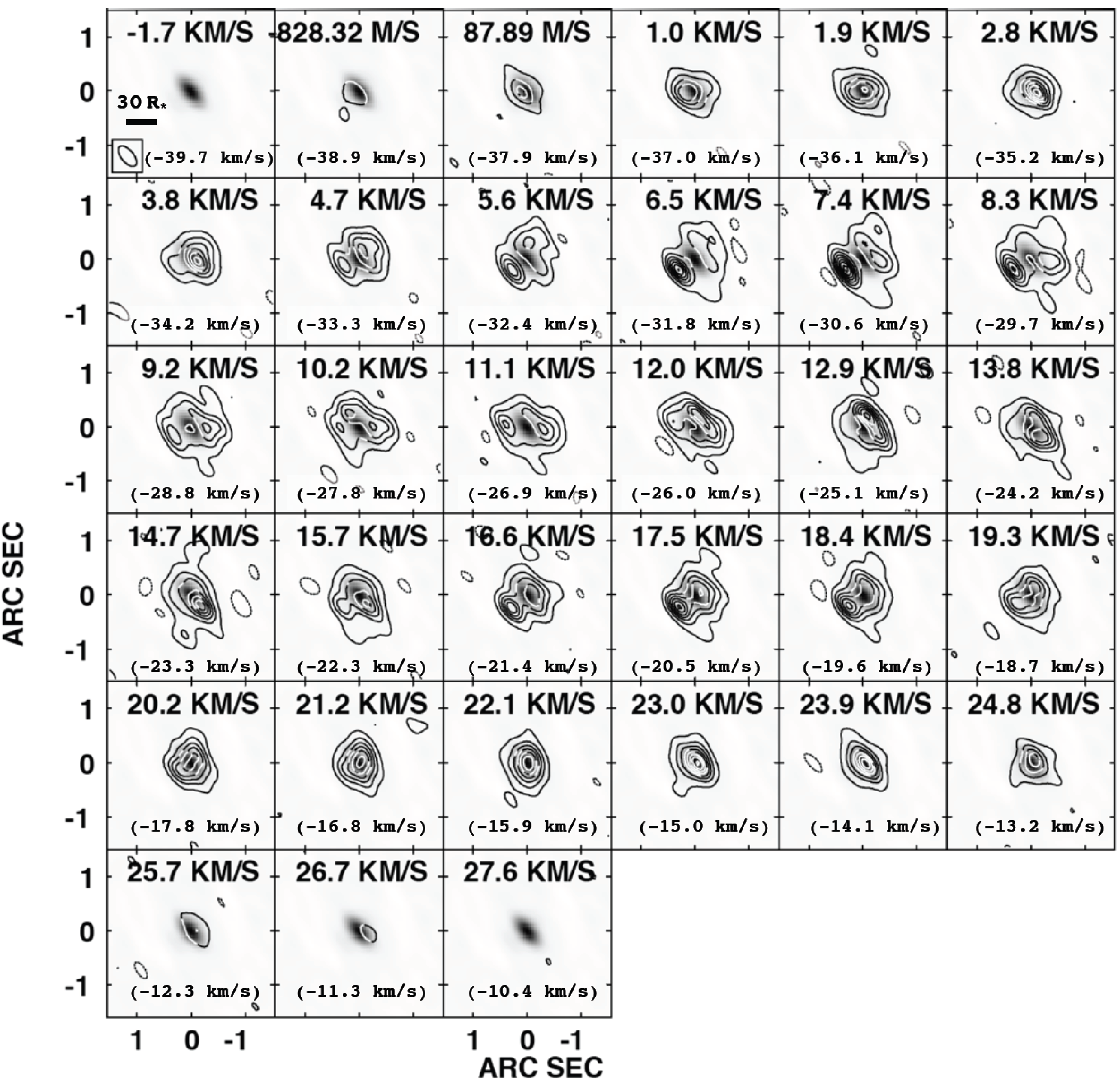}
\caption{Channel map (contours) of HCN $J=3-2$, $v$=(0, 1$^{\rm 1e}$, 0) transition, overlaid on the continuum emission of the circumstellar dust and photosphere. 
Contours are set at $-$3, 3, 9, 15, 27, 33, 39, and 45 $\sigma$ (1 $\sigma$ = 200 mJy).   
Observed \vlsr~ is shown at the top of each panel.  
The \vlsr converted to that of $v$=(0, 1$^{\rm 1e}$, 0) $J=3-2$ transition is written 
in the parenthesis at the bottom of each panel.  
The systemic velocity of the object is $-26$ \kms.  
\label{f8}}
\end{figure}

Figure \ref{f8} shows channel maps of the $v$=(0, 1$^{\rm 1e}$, 0), $J=3-2$ transition that illustrate the detailed internal structures 
of the gas 
in region II.  
Note that the systemic velocity of the object is $-26$ \kms.  
The channel maps clearly show expanding shells.   
The expansion velocity of the shells, derived by fitting the spectrum integrated over the whole region,  
is up to $\sim$ 13 \kms.  
The expanding shells are composed of many clumpy condensations.  

There are a few prominent condensations along the most copious mass-loss direction.  
By following the maps at different velocities, one can identify 
shell-like features whose radii range from 11 to 22 $R_{*}$.  
Partial shell structures are found in the channel maps at \vlsr of 3.8 and 19.3 \kms.  
The velocities are noted in the upper part of each panel.  
In other words, in these velocity ranges, the HCN maser condensations are distributed avoiding the stellar position.  
The radius of the shell is measured to be about $\sim$ 23 $R_{*}$ with a thickness of $\sim$ 15 $R_{*}$.  
They may be the smallest shell-like structures identified to date in the circumstellar envelope of this star.  

One can find one maser clump in the southeast side of the central star in the channel maps of 3.8 -- 10.2 \kms, 
noted in the upper part  
of each panel. 
Another maser clump is found in the channel maps of 
15.7 -- 19.3 \kms~ 
in the southeast side.  
Assuming 
these two clumps are located on the east side of the rim of the torus-like feature (see Section \ref{distribution}) 
and the clumps have had constant expanding velocities, 
the medium of the clumps may have departed from the central star $\ga$ 40 years ago.  
%
For the northeast side of the central star, 
a clump 
is seen in the channel maps of 4.7 -- 13.8 \kms, 
and another one is seen in the channel maps of 16.6 -- 19.3 \kms.  
Assuming a similar condition, i.e., the clumps are located on the west side of the rim of the torus-like feature 
(see Section \ref{distribution})  and 
the clump has had constant expanding velocities, 
the medium of the clumps may have been shed by the central star $\ga$ 100 years ago.  

\subsection{Acceleration of the Gas in the Innermost Circumstellar Envelope}\label{linewidth}
One can gain insights on the acceleration of the gas due to the stellar radiation from the observed line widths and the spatial distribution.   
The molecular gas is accelerated through dust by sharing their momentum through dust-gas collisions 
as the dust particles receive the radiation pressure.  
Silicon carbide (SiC) can condense into dust particle very close to the stellar photosphere at a temperature of $\sim$ 1500 K 
(e.g. Men'shchikov et al. 2001) and they may evolve chemically as going away from the star, 
while graphite grains form at lower temperature of $\sim$ 600 -- $\sim$ 1000 K 
or with radius of 5 -- 20 R$_{*}$ (e.g. Men'shchikov et al. 2001).  
At the inner edge of the envelope, SiC may play a major role to accelerate the molecular gas outwards.  

One can derive the expansion velocity by fitting line profiles.  
The line profiles were fitted assuming the envelope is spherically symmetric and expanding uniformly with the expansion velocity.  
We also assumed that the turbulence and thermal velocity is much smaller than the expanding velocity and  
the medium is ejected from the central star.   
Table \ref{tbl-1} summarizes the expansion velocities and frequency differences between recommended frequencies and 
observed central velocities measured for each line.  

The HCN $J=3-2$ transitions in the $v$=0 and (0, 1$^{\rm 1e}$, 0) states have broad line widths of $\sim$ 13 \kms,  
which is close to the terminal velocity of $\sim$ 15 \kms~ as observed from other molecular lines such as CO \citep{kna85} and 
CS \citep{you04}. 
In region II, both graphite grains and SiC concretion  
may contribute to the acceleration of the molecular gas.  
The eSMA resolved the emission in both vibrational states. 
The kinematics and distribution of the emission in the (0, 1$^{\rm 1e}$, 0) state are investigated in detail (Section \ref{distribution}).  

It is very interesting that Si$^{34}$S emission has the largest expansion velocity ($\sim$ 16 \kms) although 
the distribution is compact and is not significantly resolved (in Section \ref{distribution}). 
This expansion velocity is at the high end among the expansion velocities summarized in a detailed line survey study by Patel et al. (2008).   
Since the sublimation temperature of SiS is 1213 K \citep{han62}, which may correspond to the temperature 
in an outer layer of the stellar photosphere, 
the SiS  concretion 
mixtured with the SiS gas might receive the radiation pressure directly from the stellar radiation, 
rather than the SiS gas gains the momentum from surrounding dust particles in the region.  
The full width at half maximum of the Si$^{34}$S line is only 9.2 \kms.  
The large expansion velocity is due to the strong wing component in the redshifted side.  
It is known that the sulphur is involved in shock chemistry (e.g. Blake et al. 1987).  
It may arise from a shock region in the innermost envelope.  
The cause of the broad wing emission remains unsolved. 

On the other hand, 
the rest of the lines which have lower expansion velocities of $\lesssim$ 7 \kms, namely, 
HCN in the $v$=(1, 1$^{\rm 1f}$, 0) and (1, 1$^{\rm 1f}$, 0) states, the emission of KCl in the $v$=2 state, and 
all three unidentified lines, 
must be tracing the very innermost part of the envelope because the velocity 
becomes larger as a function of distance from the stellar photosphere \citep{gol76,win94}.  
The lowest expansion velocity is that for the HCN line in the $v$=(1, 1$^{\rm 1f}$, 0) which has 
the highest energy level among the HCN lines.  
The small frequency/velocity shifts from the catalog frequencies of the narrow line width emission (see Table 1) 
also supports the 
scenario that the expansion velocities of those components must be very small because their velocities 
are close to the systemic velocity of the star.   
Even if the molecular components of these transitions have broken shells, since the expansion 
velocities are small, their central velocities should come to very close to the systemic velocity of the star.  

\section{SUMMARY}
Using the eSMA, we investigated the properties of the innermost circumstellar envelope (CSE) of the bright carbon star, IRC+10216. 
HCN maser emission 
is spatially resolved for the first time in an astronomical object.  
KCl maser line in a vibrationally excited state is detected for the first time.  

We identified two discrete regions in the innermost circumstellar envelope.    
One is a region with a radius of less than 15 $R_{*}$ where molecular species have just formed and 
the molecular gas has just started being accelerated by stellar radiation (region I).  
This region is traced by hot HCN and KCl lines, and Si$^{34}$S line.   
Si$^{34}$S line detected in region I shows a large expansion velocity of 16 \kms with the broad wing emission unexpectedly.   
This might be partly because SiS is both in gas and solid phase and receives the radiation pressure directly from the strong radiation pressure from photosphere.  
It might be related to some kind of shock process in the innermost envelope.  
The other is a shell region with a radius of about 23 $R_{*}$ and with a thickness of $\sim$ 15 $R_{*}$ 
where the expansion velocity has reached up to 13 \kms ~(region II).  
These two discrete regions may have emerged because the size distributions of the condensed dust particles are significantly different in these regions, i.e., region II may hold larger dust particles or graphite grains.  
We identified the most copious mass loss direction using the HCN $J=3-2$ transition in the $v$=(0, 1$^{\rm 1e}$, 0)  and $v=0$ 
states. 
We propose that this direction corresponds to the equatorial direction of the star.  
The direction was found to be perpendicular to the near infrared bipolar reflection nebula and to be parallel to the features with 
a low degree of linear polarization identified at near infrared.  
The kinematics in region II are studied in detail using the HCN $J=3-2$ emission in the $v$=(0, 1$^{\rm 1e}$, 0) state,  
which 
shows clumpy structures.  
Small shell-like features with a radius of 11 -- 22 $R_{*}$ 
were identified in region II.  
A torus-like feature with a radius of $\sim$9 $R_{*}$ exists in region II.  
Assuming the expansion velocity has been constant,  the medium of the most conspicuous HCN maser clumps may have been shed by the central star 
about 40 -- 100 years ago.  

We identified nine molecular transitions in total, including hot HCN $J=3-2$ maser transitions in the 
$v$=(0, 1$^{\rm 1f}$, 1) and $v$=(1, 1$^{\rm 1f}$, 0) states 
and KCl ($v=2$) and Si$^{34}$S emission. 
These transitions may be excited directly by the strong infrared radiation from the stellar photosphere of the star and 
by line overlap of some of CO and H$_{2}$~ transitions for the HCN transition in the $v$=(1, 1$^{\rm 1f}$, 0) state.  
From the observed line profiles, expansion velocities of each line are estimated.  
All the emission lines except for Si$^{34}$S line have expansion velocities lower than the terminal velocity of the gas ($\sim$ 15 \kms) in the circumstellar envelope, which 
means that the observed emission  traces the region where the acceleration due to strong stellar radiation is taking place.  
The HCN transition in the $v$=(1, 1$^{\rm 1f}$, 0) state has the lowest expansion velocity of $\sim$ 3 \kms, suggesting that 
it arises from the innermost part of the envelope among the lines detected in this observation.  
Region II has an expansion velocity of 13 \kms.  

Future observations with eSMA will allow us to constrain physical and chemical properties of the innermost circumstellar envelope 
further in detail and construct a concrete picture of the mass loss process in a transition phase from AGB to post AGB/proto-planetary nebula (PPN).  



\acknowledgments
The Caltech Submillimeter Observatory is supported by grant number AST-0540882 from the National Science Foundation.
The eSMA developments at the JCMT are financially supported by a Netherlands NWO-M grant, the Netherlands Organization for Scientific Research, and NWO. 
The development of the eSMA has been facilitated by grant 614.061.416 from the Netherlands Organisation for Scientific Research, NWO. 
The JCMT is supported by the United Kingdom's Science and Technology Facilities Council (STFC), the National Research Council Canada
(NRC), and the Netherlands Organization for Scientific Research (NWO). 
The Submillimeter Array is a joint project between the Smithsonian Astrophysical Observatory and the Academia Sinica Institute of Astronomy and Astrophysics and is funded by the Smithsonian Institution and the Academia Sinica.
We sincerely appreciate strong support from Ray Blundell, Gary Davis and Tom Phillips, the directors of the SMA, JCMT and CSO.
HS is grateful to Ray S. Furuya, Charlie Qi, Nick Scoville, and Frank Lovas for various discussion.  
AMH is supported by a National Science Foundation Graduate Research Fellowship.



{\it Facilities:} \facility{eSMA}, \facility{CSO}, \facility{JCMT}, \facility{SMA}.

\begin{deluxetable}{lrrrrrcc}
\tabletypesize{\scriptsize}
\tablecaption{Molecular Transitions Detected with the eSMA\label{tbl-1}}
\tablewidth{0pt}
\tablehead{
\colhead{$F_{\rm cat}$\tablenotemark{a}} & \colhead{$F_{\rm cnt}$\tablenotemark{b}} & \colhead{$S_{\rm peak}$\tablenotemark{c}} & \colhead{Identification} & \colhead{E$_{L}$\tablenotemark{d}} & \colhead{$V_{\rm exp}$\tablenotemark{e}} & \colhead{$F_{\rm cat}$ $-$ $F_{\rm cnt}$} & \colhead{References for}  \\
\colhead{(MHz)}   & \colhead{(MHz)}        &\colhead{(Jy)}           & \colhead{}   & \colhead{(K)} & \colhead{(\kms)} & \colhead{(MHz)} & \colhead{Frequencies} 
}
\startdata
U264665  & ... & 0.6 & & &  4.2 (6) & ... & \\ 
264789.76  &  264790.50 & 0.8 & Si$^{34}$S $J=15-14$, $v=0$  & 88.9 &  16.3 (2) \tablenotemark{f} & $-$0.7 & (1) \\
U265253  &   ... &  0.5 &  &  & 5.1 (3) & ...& \\
265264.80 &  265265.43 & 0.8 & KCl  $J=35-34$, $v=2$ & 1012 & 4.1 (3) & $-$ 0.6 & (1) \\ 
265364.36  &  265364.90 & 1.2 & HCN  $J=3-2$, $v$=(0, 1$^{\rm 1f}$, 1) & 4045  & 6.8 (7) & $-$0.6 & (2) \\
265373.12  &  265373.99 &  0.6  & HCN  $J=3-2$, $v$=(1, 1$^{\rm 1f}$, 0) & 5773 & 3.2 (3) & $-$0.5 & (2) \\
U265386  &  ...& 0.5 &  &  & 3.5 (7) & ...& \\
265852.71  &  265852.41 & 29 & HCN  $J=3-2$, $v$=(0, 1$^{\rm 1e}$, 0) &  1036 & 12.8 (1)  & +0.3 & (1) \\
265886.19  &  265883.50 & 10 & HCN  $J=3-2$, $v$=0                        &      12.8      & 12.9 (1) \tablenotemark{g} & +2.7 & (1) \\
\enddata


\tablenotetext{a}{~Recommended frequency from catalog.}
\tablenotetext{b}{~Observed center frequency.}
\tablenotetext{c}{~Peak flux density observed with the interferometer.}
\tablenotetext{d}{~Energy of the lower level of the transition relative to the ground state.}
\tablenotetext{e}{~Expansion velocity.  The numbers in parentheses represent one standard deviation in units of the last significant digit.}
\tablenotetext{f}{~The line's full width half maximum is 9.2 \kms.  } 
\tablenotetext{g}{~The broadening effect due to hyperfine components is subtracted. }

\tablecomments{References: (1) NIST catalog; 
(2) De Lucia and Helminger 1977. 
}

\end{deluxetable}
\clearpage




\clearpage

\clearpage

\clearpage

\clearpage

\clearpage

\clearpage

\clearpage



\clearpage

\end{document}